\documentclass[prl,aps,preprintnumbers, twocolumn,showpacs, nofootinbib,superscriptaddress,notitlepage]{revtex4-1}
\usepackage{amssymb, amsthm}
\usepackage{amsmath}    
\usepackage{mathrsfs}   
\usepackage{graphicx}   
\usepackage{color}      
\usepackage{colortbl}
\usepackage{footnote}   
\usepackage{slashed}    

\usepackage[utf8]{inputenc}
\usepackage[dvipsnames]{xcolor}
\usepackage[normalem]{ulem}

\usepackage{subfiles}

\newcommand{\nn}{\nonumber}

\newcommand{\beq}{\begin{equation}}
\newcommand{\eeq}{\end{equation}}

\begin{document}


\title{Double Parton Distributions from Euclidean Lattice}

\author{Jian-Hui Zhang}
\email{zhangjianhui@cuhk.edu.cn}
\affiliation{School of Science and Engineering, The Chinese University of Hong Kong, Shenzhen 518172, China}
\affiliation{Center of Advanced Quantum Studies, Department of Physics, Beijing Normal University, Beijing 100875, China}


\begin{abstract}
We show that double parton distributions, which are important in describing double parton scattering processes in hadron collisions, can be directly computed from correlations of equal-time nonlocal Euclidean operators on the lattice in the large hadron momentum limit. 
We demonstrate this by taking the unpolarized color singlet quark double parton distribution as an example, and present a factorization formula connecting the corresponding lightcone and Euclidean correlations. This opens a new possibility of studying multiparton interactions on the Euclidean lattice.
\end{abstract}

\maketitle

\section{Introduction}
Since the deep-inelastic scattering experiment and the proposal of parton model more than 50 years ago, our understanding of the internal structure of hadrons has	 greatly improved. The well-known single parton distributions (SPDs) characterize the probability of finding a quark or gluon parton carrying certain momentum fraction $x$ inside the hadron, and play a vital role in interpreting the experimental data at high-energy colliders such as the LHC in terms of single parton scattering. While multiparton distributions, among which the double parton distributions (DPDs) are the simplest and most important ones, describe the correlated distribution of two or more partons. They can be viewed as the joint probability of finding two or more partons carrying momentum fractions $x_i$ at given relative transverse separations~\cite{Diehl:2011yj}, and enter the description of multiparton scattering processes. With the high luminosity and high energy of LHC, understanding these processes has become increasingly important in order for theoretical predictions to match the experimental accuracy and thereby optimize the potential of discovering new physics beyond the Standard Model. Such processes are also relevant in the context of heavy-ion physics~\cite{Diehl:2011yj}.

In contrast to the SPDs, the DPDs are much less known. There has been experimental effort exploring the double parton scattering processes, ranging from the study of multi-jet events at the CERN ISR and $\rm Sp\bar pS$~\cite{AxialFieldSpectrometer:1986dfj,UA2:1991apc} to the study of photon+jets at the Tevatron~\cite{CDF:1997yfa,D0:2015rpo} and double Drell-Yan-type processes at the LHC~\cite{LHCb:2016wuo,ATLAS:2018zbr,CMS:2019jcb}. Despite that, extracting the DPDs from the experimental data has been a considerably difficult task. Phenomenological analyses often need to make model assumptions such as the independence of the two partons in the DPDs~\cite{Bali:2020mij}. On the theoretical side, we do not have much guidance on the DPDs beyond the sum rules reflecting quark number and momentum conservation~\cite{Gaunt:2009re,Golec-Biernat:2014bva,Golec-Biernat:2015aza,Diehl:2020xyg} and their asymptotic behavior at small inter-parton distances~\cite{Diehl:2011tt,Diehl:2011yj,Diehl:2019rdh}. Therefore, first-principles lattice QCD can play a dominant role in providing reliable information on the DPDs, provided that they can be calculated on the lattice. However, so far only the lowest double Mellin moment of the DPDs has been calculated for the nucleon and the pion~\cite{Bali:2021gel,Bali:2020mij}.

In the past few years, significant theoretical developments~\cite{Liu:1993cv,Detmold:2005gg,Braun:2007wv,Ji:2013dva,Ji:2014gla,Ma:2017pxb,Radyushkin:2017cyf,Chambers:2017dov} have been made which allow us to directly access the Bjorken $x$-dependence of the SPDs rather than their lowest few Mellin moments from lattice QCD. Based on this, a number of lattice calculations have been carried out, yielding promising results for various SPDs (see Refs.~\cite{Ji:2020ect,Cichy:2018mum} for a recent review). These results provide valuable complementary information on the SPDs, particularly on those which are difficult to extract from experiments. 

In this work, we focus on the simplest and most important multiparton distributions, namely the DPDs, and show that they can be directly computed from correlations of equal-time nonlocal Euclidean operators on the lattice. To demonstrate this, we take the unpolarized color singlet quark DPD as an example, which is phenomenologically one of the most important DPDs, and present a factorization formula connecting the lightcone correlators defining the DPDs and the corresponding Euclidean correlators accessible on the lattice. The discussion can be easily generalized to multiparton distributions.

\section{Double parton distributions on the lightcone}
Let us begin with the unpolarized color singlet quark DPD which is defined as~\cite{Diehl:2011yj}
\begin{align}\label{eq:DPDdef}
&f_{q_1 q_2}(x_1,x_2,{y^2})=\\
&2P^+\int dy^-\int\frac{dz_1^-}{2\pi}\frac{dz_2^-}{2\pi}e^{i(x_1 z_1^-+x_2 z_2^-)P^+} h_0(y, z_1, z_2, P)\nn\\
&=2\int d\lambda\int\frac{d\lambda_1}{2\pi}\frac{d\lambda_2}{2\pi}e^{i(x_1 \lambda_1+x_2 \lambda_2)} h(\lambda, \lambda_1, \lambda_2, y^2),\nn
\end{align}
with
\begin{align}\label{eq:opdef}
&h_0(y, z_1, z_2, P)= \langle P|O_{q_1}(y,z_1)O_{q_2}(0,z_2)|P\rangle, \nn
\end{align}
\begin{align}
&h(\lambda, \lambda_1, \lambda_2, y^2)=\frac{1}{(P^+)^2}h_0(y, z_1, z_2, P),\nn\\
&O_{q}(y,z)=\bar\psi_q\big(y-\frac z 2\big)\frac {\gamma^+} {2} W\big(y-\frac z 2;y+\frac z 2\big)\psi_q\big(y+\frac z 2\big),\nn\\
&\lambda=P\cdot y, \ \ \lambda_1=P\cdot z_1, \ \ \lambda_2=P\cdot z_2,
\end{align}
where we use lightcone coordinates $v^\pm=(v^t \pm v^z)/\sqrt{2}$. $z_i^\mu=(0,z_i^-,{\vec 0}_\perp)$ are lightlike vectors, and $y^\mu=(0,y^-,{\vec y}_\perp)$ is spacelike. $|P\rangle$ denotes a hadron moving in the longitudinal direction with momentum $P^\mu=(P^+,P^-,\vec 0_\perp)$.
We take the valence combination $u-\bar u$ and $d-\bar d$ for $O_{q_1}$ and $Q_{q_2}$, respectively, in order to avoid complications due to potential mixing with gluons. 
$W\big(y+\eta_1 n; y+\eta_2 n\big)= \mathcal{P}\,{\rm exp}\big[-ig\int_{\eta_2}^{\eta_1} du\, {n}\cdot A(y+ un)\big]$ is the Wilson line along the lightcone minus direction. By doing a Fourier transform with respect to $\vec y_\perp$, the DPD in Eq.~(\ref{eq:DPDdef}) can also be turned to a DPD in momentum space~\cite{Diehl:2011yj}.

The DPD defined above involves fields at lightcone distances, and thus cannot be directly accessed on a Euclidean lattice. Nevertheless, one can define its double Mellin moments in analogy to the Mellin moments for the SPD as~\cite{Diehl:2011yj,Bali:2020mij}
\begin{align}\label{eq:doubleMM}
&\hspace{-.5em}M_{q_1q_2}^{n_1 n_2}(y^2)=\int_{-1}^1 \hspace{-.5em}dx_1 dx_2\, x_1^{n_1-1}x_2^{n_2-1}f_{q_1 q_2}(x_1,x_2, y^2)\\
&\hspace{2em}=\frac{(P^+)^{1-n_1-n_2}}{2}\int dy^- \langle P|{\cal O}_{q_1}^{+\cdots +}(y){\cal O}_{q_2}^{+\cdots +}(0)|P\rangle\nn
\end{align}
with ${\cal O}_q^{\mu_1\cdots \mu_n}$ being the twist-two quark operators from the operator product expansion (OPE)
\beq
{\cal O}_q^{\mu_1\cdots \mu_n}(y)=\bar\psi_q(y)\gamma^{\{\mu_1}i{\stackrel{\leftrightarrow}{D}{{\hspace{-.2em}}^{\mu_2}}}(y)\cdots i{\stackrel{\leftrightarrow}{D}{{\hspace{-.2em}}^{\mu_n\}}}}(y)\psi_q(y),
\eeq
where ${\stackrel{\leftrightarrow}{D}{{\hspace{-.2em}}^{\mu}}}=1/2({\stackrel{\rightarrow}{\partial}{{\hspace{-.3em}}^{\mu}}}-{\stackrel{\leftarrow}{\partial}{{\hspace{-.3em}}^{\mu}}})+igA^\mu$ is the antisymmetrized covariant derivative, and $\{\cdots\}$ denotes a symmetrization and subtraction of traces in the enclosed indices.

We can make a Lorentz covariant decomposition of the matrix elements appearing on the r.h.s. of Eq.~(\ref{eq:doubleMM})
\begin{align}\label{eq:decomp}
&\langle P|{\cal O}_{q_1}^{\mu_1\cdots \mu_{n_1}}(y){\cal O}_{q_2}^{\nu_1\cdots \nu_{n_2}}(0)|P\rangle=\\
&\hspace{3em}2P^{\mu_1}\cdots P^{\mu_{n_1}}P^{\nu_1}\cdots P^{\nu_{n_2}}\langle {\cal O}_{q_1}^{n_1}{\cal O}_{q_2}^{n_2}\rangle(\lambda, y^2)+\cdots,\nn
\end{align}
where the reduced matrix element $\langle {\cal O}_{q_1}^{n_1}{\cal O}_{q_2}^{n_2}\rangle$ is a function of Lorentz invariants $\lambda=P\cdot y$ and $y^2$ only, and the dots denote terms with uncontracted vectors $y^\mu$ and terms with the metric tensor $g^{\mu\nu}$. With Eq.~(\ref{eq:decomp}), Eq.~(\ref{eq:doubleMM}) can be turned into a manifestly covariant form~\cite{Diehl:2011yj,Bali:2021gel,Bali:2020mij}
\beq\label{eq:momrelation}
M_{q_1 q_2}^{n_1 n_2}(y^2)=\int d\lambda\, \langle {\cal O}_{q_1}^{n_1}{\cal O}_{q_2}^{n_2}\rangle(\lambda, y^2).
\eeq
In the derivation above, we have ignored the renormalization issue. The same equations can be written down for the renormalized DPDs, where the renormalization can be done following the renormalization of the SPDs~\cite{Diehl:2011yj}. This introduces the dependence on two renormalization scales. We label them as $\mu_1$ and $\mu_2$ below.

From Eq.~(\ref{eq:momrelation}), one can calculate the double Mellin moments by calculating the Lorentz invariant functions $\langle {\cal O}_{q_1}^{n_1}{\cal O}_{q_2}^{n_2}\rangle$ at  given $y^2$ and different $\lambda$, which can be realized on the lattice by evaluating the matrix elements of two currents at spatial intervals with nonzero $z$-components in a hadron at various momenta. Actually, the lowest double Mellin moment with $n_1=n_2=1$ has been calculated in this way for the nucleon and the pion in Refs.~\cite{Bali:2021gel,Bali:2020mij}. However, obtaining higher double Mellin moments is considerably difficult due to the appearance of higher-order derivative operators in the OPE. Therefore, reconstructing the DPDs from their double Mellin moments is very difficult, since this requires the knowledge of the latter to all orders.

\section{Double parton distributions from correlations of nonlocal Euclidean operators} 
In order to directly access the DPDs rather than their lower double Mellin moments on the lattice, we propose to use the following correlation of equal-time nonlocal Euclidean operators
\beq\label{eq:eucldpdME}
\tilde h(z_1,z_2, y, P)=\frac{1}{N}\langle P|O_{q_1}(y,z_1)O_{q_2}(0,z_2)|P\rangle,
\eeq
where $O_q$ has the same form as that in Eq.~(\ref{eq:opdef}) except that now both $y$ and $z_i$ are spatial vectors with $y^\mu=(0,\vec y_\perp,y^z), z_i^\mu=z_i n^\mu=(0, \vec 0_\perp,z_i)$ and $\gamma^+$ is replaced by $\gamma^z$ ($\gamma^t$ is also an option). $N$ is a normalization factor. For the Dirac matrix $\gamma^z$, we have $N=1/(n\cdot P)^2=1/(P^z)^2$. Note that here we have switched to the ordinary notation for four-vectors with $v^\mu=(v^t,\vec v_\perp,v^z)$ for convenience.

The correlation in Eq.~(\ref{eq:eucldpdME}) contains ultraviolet (UV) divergences. Since the two operators $O_{q_1}(y,z_1)$ and $O_{q_2}(0,z_2)$ are purely spatial nonlocal quark bilinear operators which are separated at a spacelike distance, their renormalization can be done in the same way as the renormalization of spatial nonlocal quark bilinear operators for the SPD~\cite{Ji:2017oey,Ishikawa:2017faj,Green:2017xeu}. In other words, one only needs to renormalize the two operators $O_{q_1}, O_{q_2}$ separately. In the discussion below, we assume that the renormalization has been done for each of the operators and converted to the $\overline{\rm MS}$ scheme with renormalization scales $\mu_1$ and $\mu_2$, respectively.

To see how the Euclidean correlation in Eq.~(\ref{eq:eucldpdME}) can be related to the DPD, we again use the OPE for each of the purely spatial nonlocal operators and write ($i=1,2$)
\begin{align}\label{eq:doubleope}
&\tilde h(z_i,\mu_i,y,P)=\!\frac{1}{4N}\!\!\sum_{n_1=1}^\infty\sum_{n_2=1}^\infty\!\!\frac{(-iz_1)^{n_1-1}}{(n_1-1)!} \frac{(-iz_2)^{n_2-1}}{(n_2-1)!}\\
&\hspace{1em}\times C_{q_1}^{(n_1-1)}(\mu_1^2 z_1^2) C_{q_2}^{(n_2-1)}(\mu_2^2 z_2^2){\tilde {\mathcal M}}_{q_1 q_2}^{n_1 n_2}(\mu_i, y,P)+\cdots,\nn
\end{align}
with
\begin{align}\label{eq:eucldecomp}
\tilde {\cal M}_{q_1 q_2}^{n_1 n_2}(\mu_i,y,P)&=n_{\mu_1}\cdots n_{\mu_{n_1}}n_{\nu_1}\cdots n_{\nu_{n_2}}\\
&\hspace{-2em}\times \langle P|{\cal O}_{q_1}^{\mu_1\cdots \mu_{n_1}}(y,\mu_1){\cal O}_{q_2}^{\nu_1\cdots \nu_{n_2}}(0,\mu_2)|P\rangle\nn\\
&\hspace{-2em}=2(n\cdot P)^{n_1+n_2}\langle {\cal O}_{q_1}^{n_1}{\cal O}_{q_2}^{n_2}\rangle(\mu_i, \lambda, y^2)+\cdots,\nn
\end{align}
where the dots represent terms with lesser powers of $n\cdot P$ and thus are suppressed in the large momentum limit. Note that the same Lorentz invariant reduced matrix element $\langle {\cal O}_{q_1}^{n_1}{\cal O}_{q_2}^{n_2}\rangle$ appears both in Eq.~(\ref{eq:decomp}) and in Eq.~(\ref{eq:eucldecomp}). Eqs. (\ref{eq:doubleope}) and (\ref{eq:eucldecomp}) indicate that $\tilde h$ is a function of the Lorentz invariants $\lambda=P\cdot y, \lambda_1=P\cdot z_1, \lambda_2=P\cdot z_2, z_1^2, z_2^2, y^2$ as well as of the renormalization scales $\mu_1, \mu_2$. Thus, we can also write $\tilde h=\tilde h(\lambda, \lambda_i,\mu_i, z_i^2, y^2)$. 

From Eq.~(\ref{eq:momrelation}), we have
\begin{align}
&\int d\lambda\, \langle {\cal O}_{q_1}^{n_1}{\cal O}_{q_2}^{n_2}\rangle(\mu_i, \lambda, y^2)=2\int_{-1}^1 dx_1 dx_2\, x_1^{n_1-1}x_2^{n_2-1}\nn\\
&\hspace{1em}\times\int d\lambda\int\frac{d\lambda'_1}{2\pi}\frac{d\lambda'_2}{2\pi}e^{i(x_1 \lambda'_1+x_2 \lambda'_2)}h(\lambda,\lambda'_i,\mu_i,y^2),
\end{align}
where we have restored the dependence on renormalization scales in $h$, as indicated by $\mu_i$.

For notational simplicity, we introduce
\begin{align}
\tilde H(\lambda_i,\mu_i,z_i^2,y^2)&=\int d\lambda\, \tilde h(\lambda, \lambda_i,\mu_i, z_i^2,y^2),\nn\\
H(\lambda_i,\mu_i,y^2)&=\int d\lambda\, h(\lambda, \lambda_i,\mu_i,y^2).
\end{align}
Using Eq.~(\ref{eq:doubleope}), we then find
\begin{align}\label{eq:tildehandh}
&\tilde H(\lambda_i,\mu_i,z_i^2,y^2)=\\
&\hspace{1em}\int_{-1}^1 dx_1 dx_2\sum_{n_1=1}^\infty\sum_{n_2=1}^\infty \frac{(-ix_1\lambda_1)^{n_1-1}}{(n_1-1)!} \frac{(-ix_2\lambda_2)^{n_2-1}}{(n_2-1)!}\nn\\
&\hspace{1em}\times C_{q_1}^{(n_1-1)}(\mu_1^2 z_1^2) C_{q_2}^{(n_2-1)}(\mu_2^2 z_2^2)\nn\\
&\hspace{1em}\times \int\frac{d\lambda'_1}{2\pi}\frac{d\lambda'_2}{2\pi}e^{i(x_1 \lambda'_1+x_2 \lambda'_2)}H(\lambda'_i,\mu_i,y^2)+\cdots.\nn
\end{align}
Define
\begin{align}
\int\frac{d\lambda_i}{2\pi}e^{iu_i\lambda_i}\!\!\sum_{n_i=1}^\infty \frac{(-i\lambda_i)^{n_i-1}}{(n_i-1)!}C_{q_i}^{(n_i-1)}(\mu_i^2 z_i^2)={\cal C}_{q_i}(u_i,\mu_i^2 z_i^2),\nn
\end{align}
where $u_i$ is in the range $[-1,1]$~\cite{Radyushkin:2017cyf}, we can then rewrite Eq.~(\ref{eq:tildehandh}) as
\begin{align}\label{eq:factcoord}
&\tilde H(\lambda_i,\mu_i,z_i^2,y^2)=\int_{-1}^1 dx_1 dx_2\int du_1\,e^{-iu_1x_1\lambda_1}\\
&\hspace{1em}\times{\cal C}_{q_1}(u_1,\mu_1^2z_1^2)\int du_2\, e^{-iu_2x_2\lambda_2}{\cal C}_{q_2}(u_2,\mu_2^2z_2^2)\nn\\
&\hspace{1em}\times \int\frac{d\lambda'_1}{2\pi}\frac{d\lambda'_2}{2\pi}e^{i(x_1 \lambda'_1+x_2 \lambda'_2)}H(\lambda'_i,\mu_i,y^2)\nn\\
&=\int du_1 du_2\, {\cal C}_{q_1}(u_1,\mu_1^2z_1^2){\cal C}_{q_2}(u_2,\mu_2^2z_2^2)H(u_i\lambda_i,\mu_i,y^2)+\cdots.\nn
\end{align}

\begin{figure}[thbp]
\includegraphics[width=.5\textwidth]{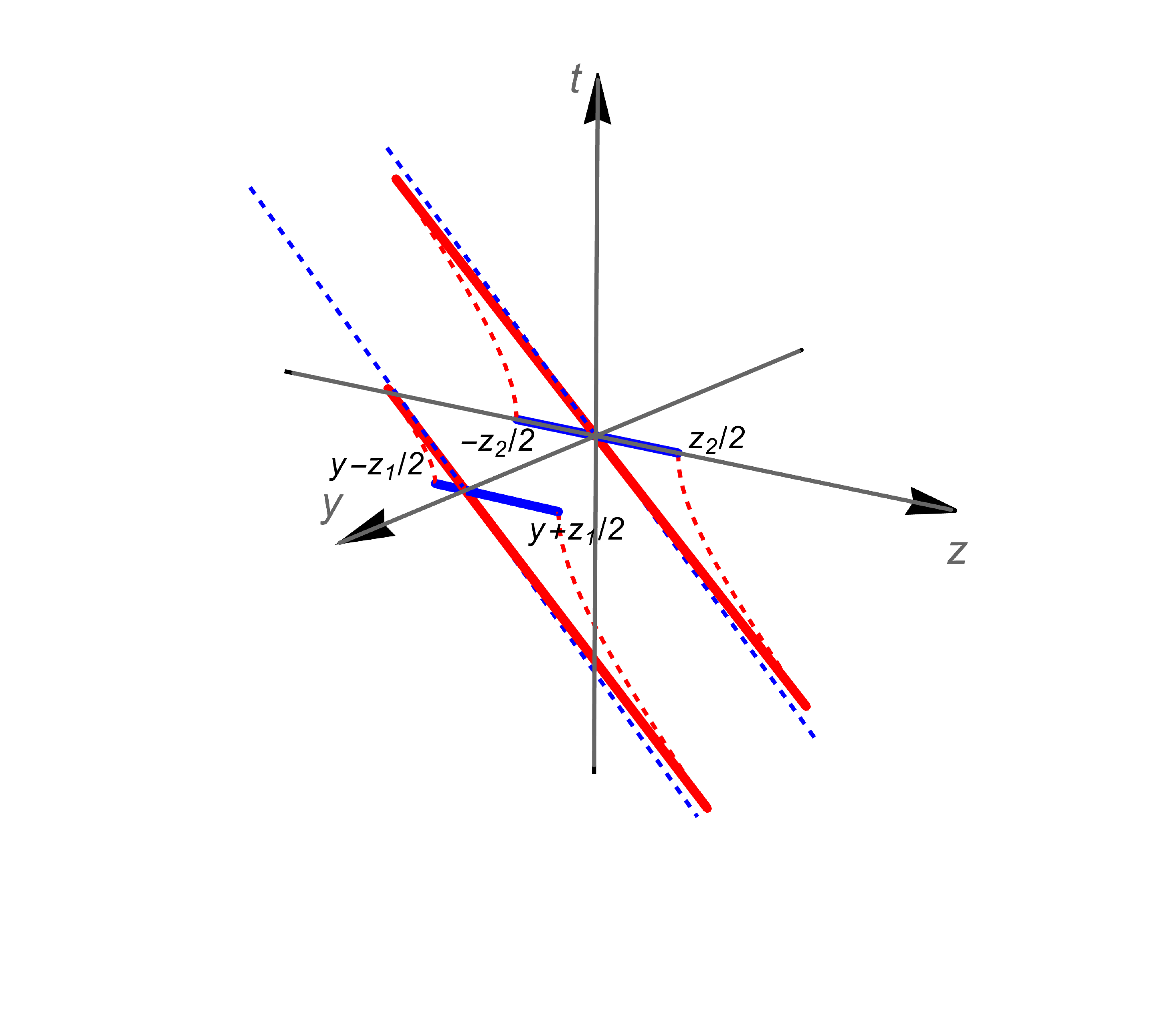}
\vspace{-6em}
\caption{Illustration of the impact of longitudinal Lorentz boost on the Euclidean equal-time nonlocal correlators (blue solid lines). The blue dashed line denotes the lightlike direction. The red solid lines represent the boosted correlators that approach the lightlike direction.}
\label{fig:dpdboost}
\end{figure}

\noindent This is the factorization formula between the correlation of equal-time nonlocal Euclidean operators and the correlation of nonlocal lightcone operators defining the DPDs, and is the main result of this work. $H$ encodes the information on the nonperturbative DPDs, while ${\cal C}_{q_i}$s are perturbatively calculable hard coefficients. From the derivation above, ${\cal C}_{q_i}$ follows from the OPE of the single quark bilinear operator, which has been calculated in Refs.~\cite{Radyushkin:2018cvn,Zhang:2018ggy,Izubuchi:2018srq,Braun:2018brg,Ji:2022thb,Chen:2020ody,Li:2020xml} and takes the following form at one-loop
\begin{align}
{\cal C}_{q_i}(u_i,\mu_i^2z_i^2)&=\delta(1-u_i)\big(1+\frac{3}{2}\frac{\alpha_s C_F}{2\pi}L_i\big)\nn\\
&-\frac{\alpha_s C_F}{2\pi}\Big(\frac{1+u_i^2}{1-u_i}\Big)_+ L_i+rem.,
\end{align}
where $L_i=\ln\Big(\frac{-\mu_i^2 z_i^2}{4e^{-2\gamma_E}}\Big)$, the subscript $+$ denotes the usual $+$-prescription, and $rem.$ represents terms depending on whether $\gamma^z$ or $\gamma^t$ is chosen for the Dirac structure in the operator.
Note that a similar coordinate space factorization formula exists for the SPDs, where the counterparts of $\tilde H$ and $H$ are often called quasi-light-front correlations and light-front correlations, respectively.

We can also perform a double Fourier transform to turn Eq.~(\ref{eq:factcoord}) into momentum space. In analogy to the SPD case, the transform can be taken either w.r.t. $\lambda_i$ with $z_i^2$ fixed or w.r.t. $z_i$ with $P$ fixed. In the former case, we have
\begin{align}
&{\cal D}(x_i,\mu_i, z_i^2,y^2)=2\int\frac{d\lambda_1}{2\pi}\frac{d\lambda_2}{2\pi}e^{i (x_1\lambda_1+x_2\lambda_2)}\tilde H(\lambda_i,\mu_i,z_i^2,y^2)\nn\\
&=\int\frac{dx'_1}{|x'_1|}\frac{dx'_2}{|x'_2|}{\cal C}_{q_1}(\frac{x_1}{x'_1},\mu_1^2z_1^2){\cal C}_{q_2}(\frac{x_2}{x'_2},\mu_2^2z_2^2)f(x'_i,\mu_i,y^2)+\cdots,
\end{align}
where ${\cal D}$ plays a similar role as that of the pseudo-distribution~\cite{Radyushkin:2017cyf} for the SPDs, and ${\cal C}_{q_i}$ has the same form as that in the coordinate space factorization Eq.~(\ref{eq:factcoord}). 

In the latter case, we have
\begin{align}
&\tilde f(x_1,x_2,\mu_i,y^2)=2\int\frac{d\lambda_1}{2\pi}\frac{d\lambda_2}{2\pi}e^{i (x_1\lambda_1+x_2\lambda_2)}\nn\\
&\hspace{1em}\times\tilde H\big(\lambda_i,\mu_i,-\frac{\lambda_i^2}{(P^z)^2},y^2\big)\nn\\
&\hspace{1em}=\int\frac{dx'_1}{|x'_1|}\frac{dx'_2}{|x'_2|}C_{q_1}\big(\frac{x_1}{x'_1},\frac{\mu_1^2}{(x'_1 P^z)^2}\big)C_{q_2}\big(\frac{x_2}{x'_2}\frac{\mu_2^2}{(x'_2 P^z)^2}\big)\nn\\
&\hspace{1em}\times f(x'_i,\mu_i^2,y^2)+\cdots,
\end{align}
where $\tilde f$ plays a similar role as the quasi-distribution~\cite{Ji:2013dva} for the SPDs, and the hard coefficients $C_{q_i}$s are related to ${\cal C}_{q_i}$s via
\begin{align}
C_{q_i}\big(\frac{x_i}{x'_i},\frac{\mu_i^2}{(x'_i P^z)^2}\big)=\int\frac{d\lambda'_i du_i}{2\pi}  \, e^{i(\frac{x_i}{x'_i}-u_i)\lambda'_i}{\cal C}_{q_i}\big(u_i,-\frac{\mu_i^2{\lambda'_i}^2}{(x_i' P^z)^2}\big).\nn
\end{align}
One can also perform a Fourier transform w.r.t. $\vec y_\perp$ on the above results to convert them to the conjugate momentum $\vec\Delta_\perp$ space. Note that in this case an inhomogeneous term in the hard coefficients might be generated, in analogy to that appears in the double evolution equation for the renormalized DPDs in $\vec\Delta_\perp$ space~\cite{Diehl:2011yj}.

To illustrate the connection between the nonlocal Euclidean correlators and the lightcone correlators in the factorization formula, we depict in Fig.~\ref{fig:dpdboost} the impact of a longitudinal Lorentz boost on equal-time nonlocal Euclidean correlators. As can be seen from the figure, under a large Lorentz boost both Euclidean correlators approach the lightcone ones. Although they cannot be literally on the lightcone, their difference from the lightcone correlators is perturbatively calculable in the large momentum limit, thanks to the asymptotic freedom of QCD. 

The approach presented here provides a general recipe for directly accessing double parton physics from Euclidean lattice. It can be extended to the case where the transverse momentum of each parton is included or when the kinematics becomes non-forward. Actually, some preliminary one-loop study~\cite{scettalk} has indicated that the infrared contributions are the same for the Euclidean and lightcone correlators defining the transverse-momentum-dependent DPDs. Moreover, it can be generalized to studying general multiparton physics. We will investigate these issues in future publications~\cite{zhangmpp}.


\section{Summary and outlook}
To summarize, we have presented an approach that allows us to directly compute the DPDs rather than their lower double Mellin moments from Euclidean lattice. We demonstrate it by taking the unpolarized color singlet quark DPD as an example, and derive a factorization formula connecting the Euclidean correlators and the lightcone ones defining the DPD. The approach has the potential to be applied to studying general multiparton physics. It will be interesting to see to what extent it can help with our understanding of the multipartonic structure of hadrons.

{\it Note}: After finishing this paper, we learned that considerations for $k_t$-dependent DPDs along similar lines have been made by the Amsterdam group~\cite{scettalk}.

\begin{acknowledgments}
\section*{Acknowledgments}
We thank Xiangdong Ji and Andreas Sch\"afer for valuable discussions and comments. This work was supported in part by the National Natural Science Foundation of China under Grant No. 11975051 and a NSFC-DFG joint grant under grant No. 12061131006 and SCHA~458/22.
\end{acknowledgments}

\bibliographystyle{apsrev}
\bibliography{ref}

\clearpage


\end{document}